# The critical relaxation of the model of iron-vanadium magnetic superlattice

Vadim A. Mutailamov [1], Akai K. Murtazaev [1,2]

[1] Institute of Physics DSC RAS, 94 M.Yaragskii Str., Makhachkala, Russia, 367003.
E-mail: vadim.mut@mail.ru

[2] Daghestan State University, 43a M.Gajiev Str., Makhachkala, Russia, 367025.
E-mail: akai2005@mail.ru

ABSTRACT

The critical relaxation of iron-vanadium magnetic superlattice in case of the equality between interlayer and intralayer exchange interactions is investigated. The dynamic and static critical exponents of the model are calculated. A value of the critical temperature is evaluated.

## 1. INTRODUCTION

Researches on metallic nonmagnetic superalttices consisting of alternate atomic layers of magnetic and non-magnetic materials are of great interest in the modern condensed-matter physics [1-3]. The possibility to control the fundamental properties of superlattices (magnetization, interlayer exchange interaction, magnetoresistance, and other characteristics) by means of external action allows creating structures with predetermined parameters, what makes these materials unique objects for the practical application and theoretical investigation.

Since the experimental researches of such systems meet with essential difficulties, the computational physics methods came to be successfully used for their study recently. In works [4-6], the static critical behavior of magnetic $Fe_2/V_{13}$ superlattices is investigated, static critical exponents are calculated, and their dependence on correlation of intralayer and interlayer exchange interactions is studied. The critical exponents of studied superlattice models are established to be dependent on a value of interlayer exchange interaction parameter. At the same time the scaling correlations between critical exponents are carried out with extremely high precision. This situation deviates from the modern theory of phase transitions and critical phenomena. In this regard, the investigation of critical dynamics of these models arouses great interest that can be a principal for an explanation of difficulties appearing in the exploration of static critical phenomena.

Recently, the critical dynamics of magnetic materials models are successfully studied using a short-time dynamic method [7-9], where the critical relaxation of a magnetic model from non-equilibrium state into the equilibrium one is investigated within *A* model (Halperin and Hohenberg classification of universality classes of dynamic critical behavior [10]). Traditionally, it is considered that a universal scaling behavior exists in the state of thermodynamic equilibrium only. Nevertheless, a universal scaling behavior for some dynamic systems is shown to be realized at earlier stages of their time evolution from high-temperature disordered state into the state corresponding to the phase transition temperature [11]. Such a behavior is realized after a certain time period which is rather larger in a microscopic sense, but remains small macroscopically. The same picture is observed when the system evolving from low-temperature ordered state [7-8].

## 2. INVESTIGATION METHOD

Using the renormalization group method, the authors [11] showed that far from equilibrium point after microscopically small time period, a scaling form is realized for *k*-th moment of the magnetization

$$M^{(k)}(t,\tau,L,m_0) = b^{-k\beta/\nu} M^{(k)}(b^{-z}t, b^{1/\nu}\tau, b^{-1}L, b^{x_0}m_0), \tag{1}$$

where $M^{(k)}$ is *k*-th moment of the magnetization; *t* is the time; $\tau$ marks the reduced temperature; *L* denotes the linear size of the system; *b* is a scale coefficient; $\beta$ and $\nu$ are static critical exponents of magnetization and correlation radius; *z* denotes a dynamic critical exponent; $x_0$ denotes new independent critical exponent defining the scaling dimension of the initial magnetization $m_0$.

When starting from low-temperature ordered state ($m_0 = 1$) in the critical point ($\tau = 0$), assuming $b = t^{1/z}$ in Eq.(1), for the systems with sufficiently large linear sizes the theory predicts an power-law behavior of the magnetization in the short-time region

$$M(t) \sim t^{-c_1}, \quad c_1 = \frac{\beta}{\nu z}. \tag{2}$$

Finding the logarithm of both parts of Eq. (2) and taking derivatives with respect to $\tau$ at $\tau = 0$ we get the power law for the logarithmic derivative

$$\partial_\tau \ln M(t,\tau)\big|_{\tau=0} \sim t^{-c_{l1}}, \quad c_{l1} = \frac{1}{\nu z}. \tag{3}$$

For Binder cummulant calculated by first and second moments of magnetization, the finite-size scaling theory gives dependence at $\tau = 0$:

$$U_L(t) = \frac{M^{(2)}}{(M)^2} - 1 \sim t^{c_U}, \quad c_U = \frac{d}{z}. \tag{4}$$

So, during one numerical experiment, the short-time dynamics method allows to determine the values of three critical exponents $\beta$, $\nu$, and *z* using correlations (2-4). Moreover, the dependences

(2) plotted at different temperature values permits to detect a value $T_c$ by their deviation from direct line in the log-log scale.

### 3. MODEL

We study the critical relaxation from low-temperature ordered state of $Fe_2/V_{13}$ superlattice by means of the short-time dynamics method. In microscopic model of the superlattice offered in works [4-6], every atom of iron has four nearest neighbors from an adjacent iron layer. The iron layers are shifted relative one another on half lattice constant along the *x* and *y* axes. The magnetic moments of iron atoms are ordered in *xy* plane. The scheme of iron-vanadium sublattice is shown in Fig.1.

An interaction between the nearest neighbors within layer has a ferromagnetic character and is determined by the exchange interaction parameter $J_\parallel$. The interlayer interaction $J_\perp$ between vanadium magnetic layers is transferred by the conduction electrons in the non-magnetic interlayer of vanadium (RKKI-interaction). In the real sublattices, its value and sign can change depending on a number of adsorbed hydrogen into the vanadium subsystem. Since an accurate dependence of the RKKI-interaction is unknown, usually, when carrying out the numerical investigations the whole range of interlayer interaction values from $J_\perp = -J_\parallel$ to $J_\perp = J_\parallel$ is studied. In this work, we present one special case of RKKI-interaction, when $J_\perp = J_\parallel$.

As in the experiment the distance between magnetic layers is substantially larger than the interatomic distance, every atom interacts with the averaged moment of neighbor layers. A size of averaging region is a model parameter. Our investigations are made for limiting case, when every atom interacts with only one nearest atom from neighbor layer. The study of the static critical behavior of magnetic superlattices [4-6] showed that this approach describes a critical behavior of these models to the best advantage.

Thus, the Hamiltonian of this model can be written as modified 3D XY-model [4-6]

$$H = -J_\parallel \frac{1}{2} \sum_{i,j} \left( S_i^x S_j^x + S_i^y S_j^y \right) - J_\perp \frac{1}{2} \sum_{i,k} \left( S_i^x S_k^x + S_i^y S_k^y \right), \qquad (5)$$

where first sum takes into account the direct exchange interaction of each magnetic atom with nearest neighbors inside the layer, and second denotes the RKKI-interaction with atoms of neighboring layers through non-magnetic interlayer; $S_i^{x,y}$ marks the components of the spin localized in a site *i*.

## 4. RESULTS

We study a system with the linear size $L=64$ containing 262144 spins in a case of equality between intralayer and interlayer exchange interactions. Let`s note that in this case, according to Eq. (5), the Hamiltonian of studied model is similar to the Hamiltonian of the classical 3D XY-model. The investigations are carried out by Metropolis standard algorithm of Monte-Carlo method. The relaxation of the system is performed from initial fully ordered state with starting value of magnetization $m_0=1$ during $t_{max}=1000$, where one Monte-Carlo step per spin is taken as "time" unite. Relaxation dependencies are calculated up to 14 000 times, obtained data are averaged among themselves.

The critical temperatures are defined by the dependence of the magnetization on the time Eq. (2), which, in a point of phase transition, must be a straight line in log-log scale. The deviation of the straight line is estimated by the least-squares method. The temperature, at which this deviation is minimal, is taken as a critical. Fig. 2 presents the time dependence of magnetization at three values of temperature around the phase transition point in log-log scale (here and further all values are presented in arbitrary units). Estimated value of the critical temperature in units of exchange integral $k_bT/J$ is $T_c=1.752(1)$. The logarithmic derivative in a phase transition point is calculated by the least-squares approximation by three time dependences of magnetization plotted at temperatures $T=1.742$, $T=1.752$ and $T=1.762$.

In Fig. 3, the time dependence of Binder cummulants $U_L$ is demonstrated in log-log scale at phase transition temperature. The analysis of curve shows that exponential scaling behavior $U_L(t)$ is realized from a time point $t=100$.

Therefore, the overall least–squares approximation by Eq. (4) is carried out within time ranges $t=[100;1000]$. This had resulted in exponent value $C_U=1.54(3)$. According to Eq. (4) the overall value of dynamic critical exponent is $z=1.95(3)$, what is close to the theoretical value predicted for anisotropic magnets ($z=2$, model A [10]).

The dependences of magnetization and magnetization derivative on the time in log-log scale are presented in Fig. 4 and 5 respectively. The data are approximated within time ranges $t=[100;1000]$ like Binder cummulants. The exponent values $c_I=0.29(3)$ and $c_{II}=0.79(3)$ have resulted from approximation. The critical exponents of magnetization $\beta=0.36(3)$ and correlation radius $\nu=0.65(3)$ are calculated by means of Eqs. (2-4).

In works [4-6], where the static critical behavior of $Fe_2/V_{13}$ superlattice model was studied by traditional equilibrium method, $\beta=0.342(3)$ and $\nu=0.671(3)$ for the similar exponents. The theory predicts $\beta=0.3485(3)$ and $\nu=0.67155(37)$ for classical XY-model [12]. As it is evident, our values of static critical exponents agree with results of both works.

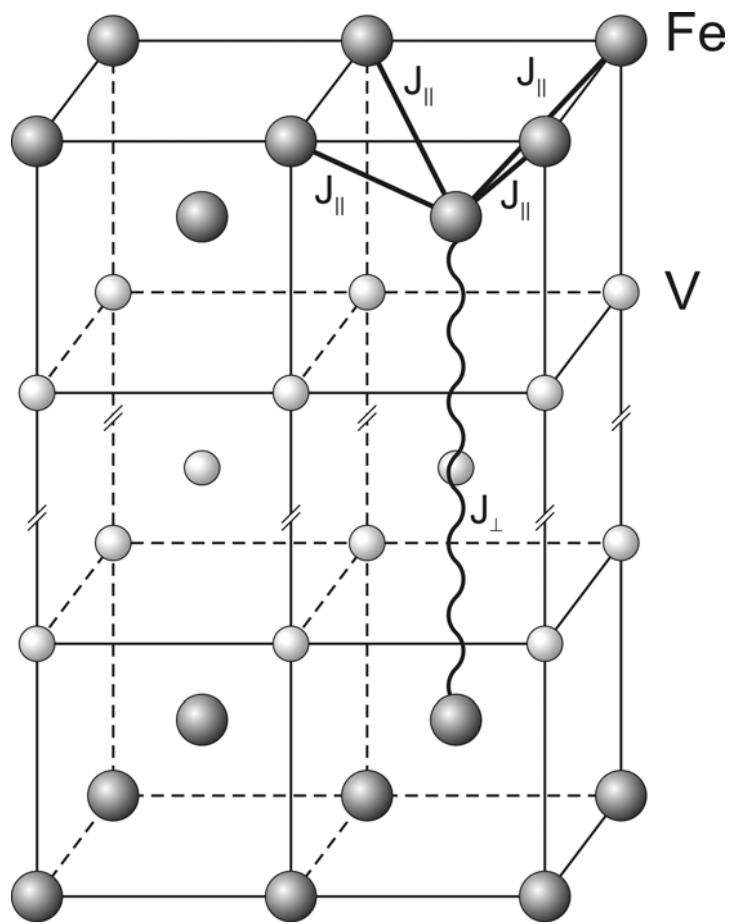

Fig.1. The scheme of iron-vanadium Fe$_2$/V13 superlattice. Three of the thirteen vanadium monolayers presented for the clarity.

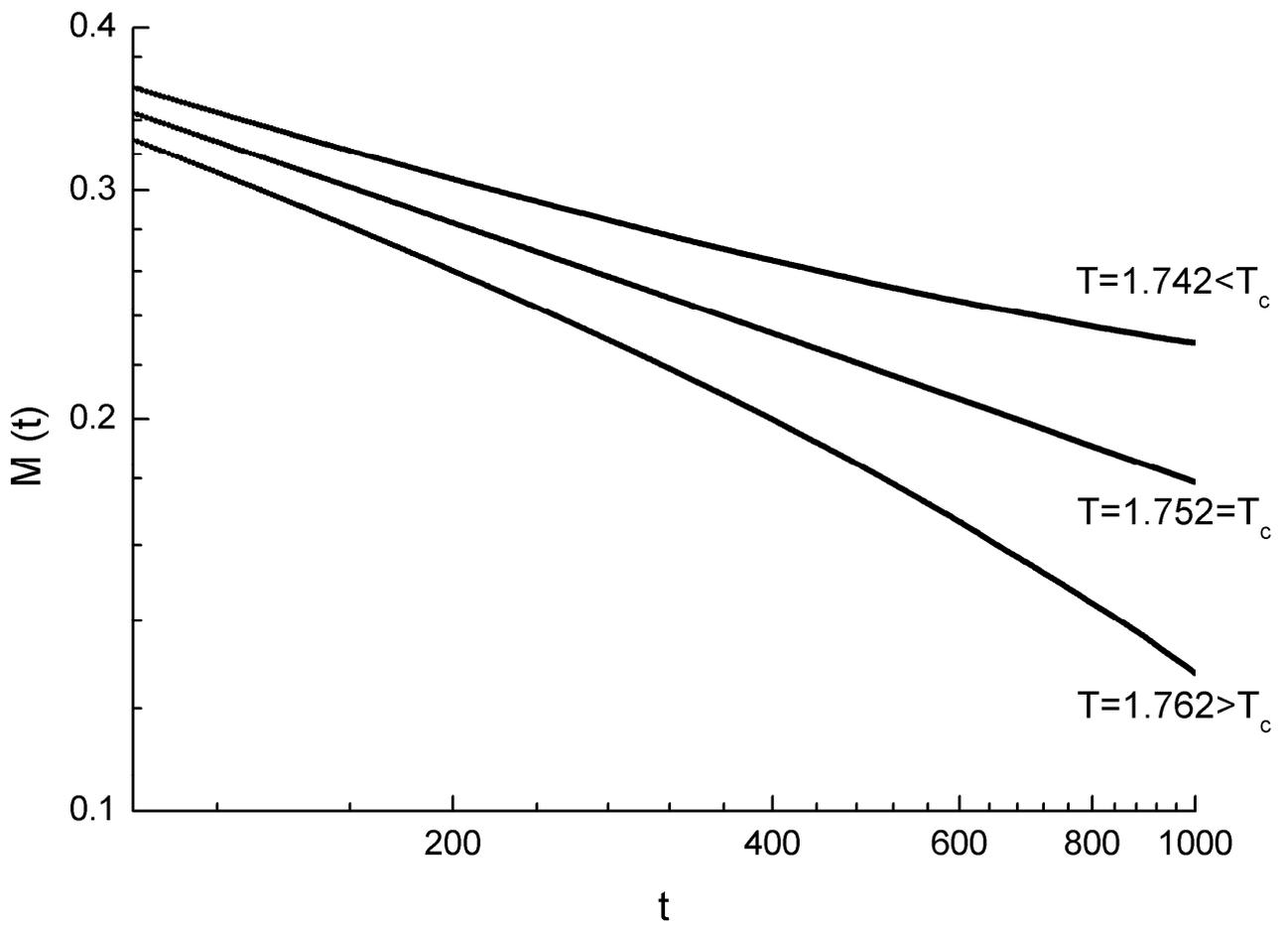

Fig.2. Time evolution of the magnetization at three different temperatures around the phase transition point.

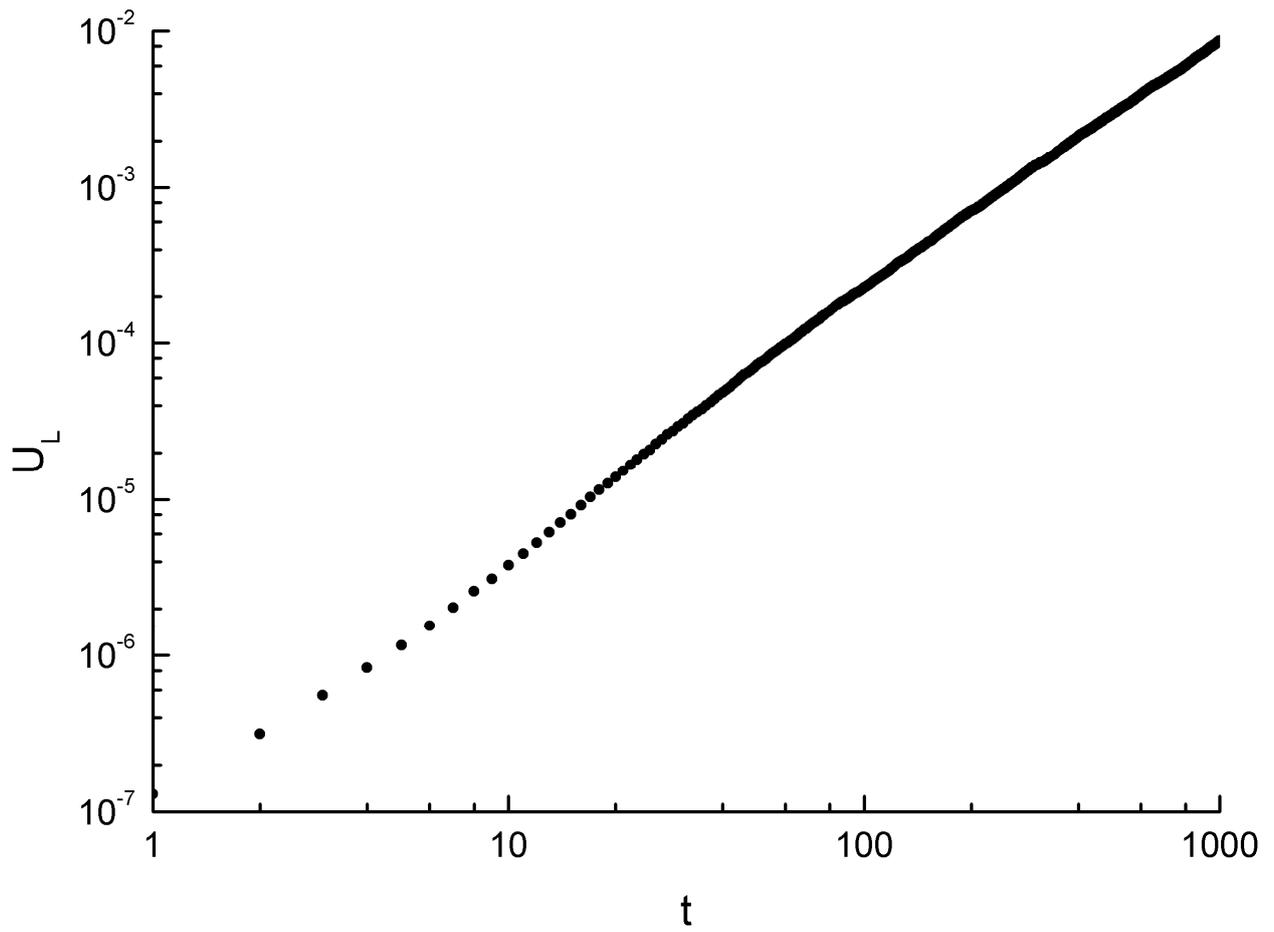

Fig.3. Time evolution of the Binder cummulants at the phase transition temperature.

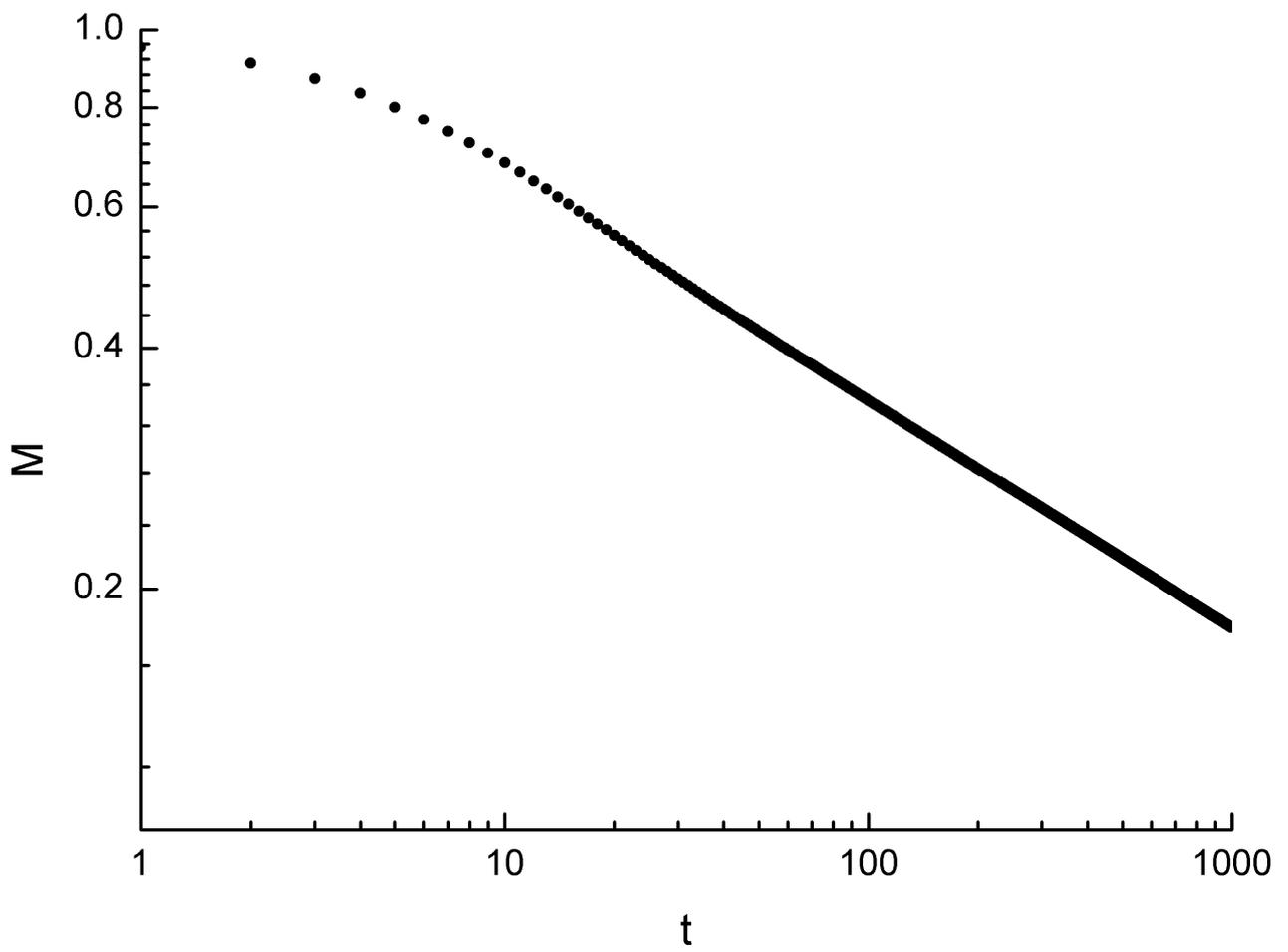

Fig.4. Time evolution of the magnetization at the phase transition temperature.

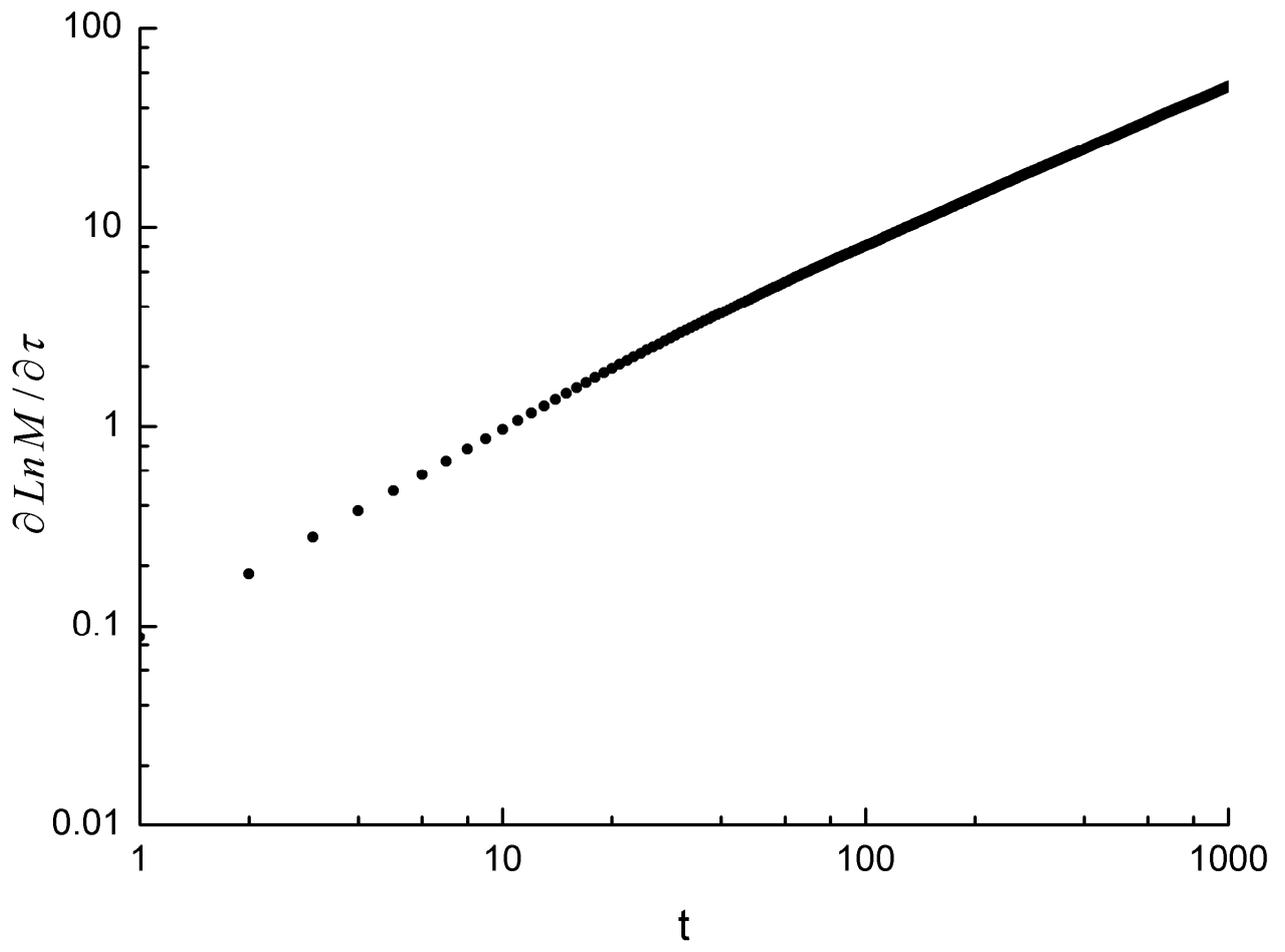

Fig.5. Time evolution of the magnetization derivative at the phase transition temperature.